\documentclass[sigconf]{acmart}
\usepackage{booktabs}
\setcopyright{rightsretained}

\usepackage{amsfonts,amssymb,amstext,latexsym}
\usepackage{multirow}
\usepackage{caption}
\usepackage{subfigure}
\usepackage{graphicx}
\usepackage[ruled, noend, vlined, linesnumbered]{algorithm2e}
\SetKwInput{KwInput}{Input}
\usepackage{url}
\definecolor{darkgreen}{rgb}{0,0.5,0}
\usepackage{xargs}
\everypar\expandafter{\the\everypar\looseness=-1 }
\let\OLDthebibliography\thebibliography
\renewcommand\thebibliography[1]{
  \OLDthebibliography{#1}
  \setlength{\parskip}{0pt}
  \setlength{\itemsep}{0pt plus 0.3ex}}
\usepackage[colorinlistoftodos,prependcaption,textsize=tiny]{todonotes}
\newcommandx{\question}[2][1=]{\todo[linecolor=pink,backgroundcolor=pink!25,bordercolor=black,#1]{#2}}
\newcommandx{\unsure}[2][1=]{\todo[linecolor=red,backgroundcolor=red!25,bordercolor=red,#1]{#2}}
\newcommandx{\change}[2][1=]{\todo[linecolor=blue,backgroundcolor=blue!25,bordercolor=blue,#1]{#2}}
\newcommandx{\info}[2][1=]{\todo [linecolor=green,backgroundcolor=green!25,bordercolor=green,#1]{#2}}

\usepackage{xspace}

\newcommand{\card}[1]{\rho(#1)}
\newcommand{\BC}[1]{BC(#1)}

\renewcommand\footnotetextcopyrightpermission[1]{}
\begin{document}

\title{Active Betweenness Cardinality: Algorithms and Applications}
\author{Yusuf Ozkaya}
\affiliation{School of Computational Science and Engineering \\
\institution{Georgia Instuitute of Technnology}}
\email{myozka@gatech.edu}

\author{A. Erdem Sariyuce}
\affiliation{\institution{
Computer Science and Engineering \\
University at Buffalo} }
\email{erdem@buffalo.edu}

\author{\"Umit V. \c{C}ataly\"urek}
\affiliation{School of Computational Science and Engineering \\
\institution{Georgia Instuitute of Technnology}}
\email{umit@gatech.edu}

\author{Ali Pinar}
\authornote{Supported by the Laboratory Directed Research and Development program at Sandia National Laboratories.
Sandia National Laboratories is a multimission laboratory managed and
operated by National Technology and Engineering Solutions of Sandia,
LLC., a wholly owned subsidiary of Honeywell International, Inc., for
the U.S. Department of Energy's National Nuclear Security Administration
under contract DE-NA-0003525.}
\affiliation{   \institution{Sandia National Laboratories\\ Livermore, CA}}
\email{apinar@sandia.gov}

\begin{abstract}
Centrality rankings such as degree, closeness, betweenness, Katz, PageRank,
etc. are commonly used to identify critical nodes in a graph. These methods
are based on two assumptions that restrict their wider applicability. First,
they assume the exact topology of the network is available. Secondly, they do
not take into account the activity over the network and only rely on its
topology. However, in many applications, the network is
autonomous, vast, and distributed, and it is hard to collect the exact
topology. At the same time, the underlying pairwise activity  between node
pairs is not  uniform and node criticality strongly depends on the
 activity on the underlying network.

In this paper, we propose  {\em active betweenness cardinality}, as a new
measure,  where the node criticalities are based on not the static structure,
but the activity of the network.
We show how this metric can be computed efficiently by using only local
information for a given node and  how we can find the most critical nodes
starting from only a few nodes. We also show how this metric  can be used to
monitor a network and identify failed nodes. We present experimental results
to show effectiveness  by demonstrating how the failed nodes can be identified
by measuring active betweenness cardinality of a few nodes in the system.
\end{abstract}

\keywords{graph  mining;  centrality measures; critical node detection; anomaly detection; change detection; }

\maketitle

\section{Introduction}\label{sec:intro}

Identifying critical elements of a network has been at the heart of many
efforts on infrastructure security and social network analysis.  The critical
elements are  those, whose presence are essential for a network to maintain its
functionality at or near its maximum. For an infrastructure network, a critical
node may be a major hub, for an airline network, or a router in a communication
network. In a social network, a critical element can be an influential person.
Subsequently, the network science community have proposed many metrics to
quantify the criticality of network elements and associated algorithms to
compute these metrics efficiently.

A major thrust in this area has been {\em  network centrality}, which assigns a
number that is a measure for its importance to each node. The importance of a
node varies from one application to another, and thus there have been many
metrics to measure centrality.  For instance, {\em Closeness}~\cite{CC1} and
{\em harmonic} centrality~\cite{CC2}  of a node are based on the distance from
this node to all the other nodes. The smaller the distances are, the higher
will be the centrality of the node. { \em Betweenness}  centrality~\cite{BC} of
a node on the other hand, is based on how often this node appears on a shortest
path between any pair of nodes.   The more often this node is on a shortest
path, the higher will be its betweenness centrality. One can also consider
degree centrality, which looks at the number of direct neighbors of a node, or
its generalization, Katz centrality~\cite{KC}, which takes into account not
only 1-hop neighbors as in degree centrality, but all 2, 3, ... hop
neighborhoods  but decreases the  contribution of each node  based on the
distance. For more detailed information of centrality measures, we refer the
reader to~\cite{SariyuceJPDC}.  In this paper, we will work on a variant of
betweenness centrality as we discuss below.

These measures  provide a wide variety of valuable insight into  graphs, but
they all rely on two assumptions:  (1) centrality of a node depends only on
the topology of a graph and does not take into account the dynamics on the graph
or how the graph is being utilized, and  (2) we  know the exact topology of the
whole graph. These assumptions restrict the effectiveness and applicability of
these methods. First, we cannot assume there will be an interaction between all
pairs of nodes on a large graph at the same level or frequency. This may be due
to lack of interest (e.g., not every user visits every page on the Web) or some
nodes provide identical services and users only interact with the nearest such
server (one goes to a close-by hospital, not necessarily all the hospitals).
One question we could ask is that can we develop approaches with models that
comprise probabilities of  interactions between every pair of vertices? While
this sounds good at first, building such a vast model with high confidence can
be as burdensome as  computing the centralities themselves. It should be noted
that  centrality measures were initiated by the social science studies on much
smaller networks, where neither of the these two limitations apply. But one
should be careful before applying the same ideas to current datasets, where the
graphs are much larger  and  comprise nodes with identical functionalities.

The second problem is the assumption that we know the exact topology of the
network. Many networks such as the Internet are vast, distributed, and
autonomous. Moreover, both their topology and the dynamics on them keep
evolving. Determining their exact topology or even predicting their topological
features are far from trivial.

Here we propose a new approach for centrality that takes into account the
network dynamics and does not require having access to the exact topology. We
call our approach the {\em Active Betweenness Cardinality (ABC)} model. The
metric for our model is the number of distinct pairs of nodes that communicate
through this node within a specified period. Our data access model is that for
any interaction through a node, we know the source and the destination.  For a
communication network, this would be the source IP and the destination IP; for
an airline network, this would be departure and destination locations for each
passenger.  Note that, we can compute the ABC value of any node with only local
computation on that node without knowing anything about the rest of the graph.
As we will show in this paper, we can estimate this quantity efficiently (both
in terms of runtime and memory) and accurately, using a sublinear algorithm. We
will also show that  we can search and identify  nodes with high ABC values.
Moreover, we will show that, the proposed metric can help us identify presence
and location of changes in a network. We claim that this approach  provides a
critical capability for analysis for vast distributed networks.

\textit{Outline.} The rest of the paper is organized as follows:
Section~\ref{sec:prel} introduces the preliminaries on
estimating set cardinalities  and centrality on graphs. Section~\ref{sec:exp_setup}
presents our experimental setup.
The proposed metric is explained in Section~\ref{sec:acc}, followed by techniques that can approximate this metric and  finding critical nodes starting from a small number of nodes.
Section~\ref{sec:detecting_changes} presents how the proposed techniques can be used to detect  significant changes in the network.
Section~\ref{sec:conc} concludes the paper.

\subsection{Our contributions: }
This paper  presents significant contributions on several fronts.

\begin{itemize}

\item {\em A novel metric for criticality.}  We propose { \em Active
Betweenness Cardinality (ABC)} metric to identify critical nodes in a network.
This metric is based on the  number of distinct pair of nodes that communicate
through this node within a specified period.

\item {\em A sublinear algorithm to compute this metric.}  Proposed metric
requires estimating the cardinality of  a set, whose elements we observe with
repetition. We use  the hyperloglog algorithm to show that  the proposed metric
can be computed with a small amount of memory, making the proposed approach
attractive for  massive networks.

\item {\em Algorithms to find nodes that maximize this metric.} Previous
techniques can estimate the centrality of a node. Can we find nodes with the
highest centrality values without exhaustive search? We designed an algorithm
that finds high-centrality nodes by walking from a specified subset of nodes.

\item  {\em Failure detection.}  We define critical elements  as those nodes
whose failure will affect the remainder of the system. Symmetrically, these
elements are affected by failure of other important components in the system.
We show that out metric can be used to identify presence and location of
failures of other network elements.

\item {\em Thorough experimental analysis.} We show that  the effectiveness of
all methods through detailed  empirical studies. We show that the proposed metric can be estimated
efficiently using only local algorithms and show how this algorithms can be used to search for the most critical nodes.
We also show that the new metric can be used to detect failed nodes by suing classification methods.
\end{itemize}

\section{Preliminaries}
\label{sec:prel}

Let $G = (V,E)$ be a simple undirected, unweighted graph where $n$ is the number of nodes and $m$ is the number of edges.
We assume that any pair of nodes can interact with each other by
sending/receiving some information and any interaction can be repeated. We
define $G$ as an \textbf{active network}. If there is an interaction between
nodes $u$ and $v$, it happens via a path $p$ between $u$ and
$v$, and all the communication thereafter also happens via the same path $p$.
For consistency with  the real-world routing
algorithms~\cite{RFC2453}, we restrict $p$ to be one of the shortest paths
between $u$ and $v$. We define the number of \textbf{unique}
interactions that are transmitted by a node $u$ as the \textit{cardinality} of
$u$, denoted as $\card{u}$. We want to detect the nodes with large cardinality
by using only the local information.

\subsection{Cardinality Estimation on Data Streams }
\label{sec:HLL}

An important part of our problem is about estimating the cardinality of a data
stream, which is also called the count-distinct problem. We observe the
elements of a set in a stream which can occur repeatedly, not just once. Our
goal is to estimate the cardinality of this set.

Formally, given a stream $x_0,x_1, x_2, \ldots x_K$ with repetitions, where
$x_i\in S$ for $i=0,\ldots K$. Note that due to repetitions $x_i$ and $x_j$ are
not necessarily distinct. We have a dual objective function: we want to
estimate $|S|$, the cardinality of the set $S$, as accurately as possible by
using minimal storage. For the purposes of this paper, the elements of the
stream will be the interacting pairs, and each entry in the system will be one
message exchange.

The difficulty of this problem lies in handling the repetitions. A trivial
solution is to maintain a hash table to keep track of the previously observed
elements. This can give an exact solution, but associated memory requirement
will be linear in the size of the set $S$, which is impractical for many real-
world scenarios. The algorithmic challenge is in drastically reducing storage,
while maintaining accuracy. Randomized algorithms are helpful to address this
challenge. Using randomization for the cardinality estimation was initiated by
Flajolet and Martin~\cite{FlMa85}. Since then, it has been the subject of many
research efforts. A thorough survey of this literature is beyond the scope of
this paper and we refer to readers to~\cite{HeNu13} and references therein. Our
methods do not depend on a particular cardinality estimation method, but we
briefly describe the algorithms used in our experiments to show that
cardinalities on data streams can be estimated accurately and efficiently.

The Hyperloglog (HLL) algorithm for cardinality estimation was proposed by
Flajolet et al.~\cite{FlFu07} as an improvement over the Loglog
algorithm~\cite{DuFl03}, and it is the best algorithm, both in theory and in
practice, for estimating large cardinalities~\cite{HeNu13}. Hyperloglog
algorithm is based on a hash function that can transform stream elements into
uniformly distributed random numbers. The key observation is that the
cardinality of a multiset of uniformly distributed random numbers can be
estimated by the maximum number of leading zeros in their binary
representations -- if the maximum number of leading zeros observed is $n$, then
the cardinality of the set can be estimated as $2^n$. However, a direct
application of this idea suffers from a large variance. To solve this, the
input is divided into $m=2^p$ separate buckets based on the leading $p$ bits of
each number to estimate the cardinality of each bucket and then these
individual estimates are combined using harmonic mean to get an estimate for
the full set. As a  practical guide on the performance, the Hyperloglog
algorithm can compute estimates with relative  error of $1.04/\sqrt{m}$  and
requires $O(\epsilon^{-2}
\log\log n+\log n)$ space to provide  an $(1\pm \epsilon)$ -approximation with a
high probability of success.
In our  experiments we used the Hyperloglog method with parameter $p = 12$
bits, and thus, $m = 2^p = 2^{12}$ buckets.

\subsection{Centrality Measures}
\label{sec:cent}

The centrality metrics play an important role in network and graph analysis
since they are related with several concepts such as reachability, importance,
influence, and power~\cite{simsekb08,Jin10,LeMerrer2009,Pham10,Shi11}.
Betweenness and closeness centralities~(BC and CC) are two such metrics.
However, the complexity of the best algorithms to compute them are unbearable
for today's large-scale networks: it is $\mathcal{O}(nm)$ for unweighted
networks~\cite{brandes2001}. This already makes the problem hard even for
medium-scale graphs, and thus many research efforts have focused on efficiently
utilizing state of the art HPC platforms for this
problem~\cite{Jia11,Sariyuce13-GPGPU,SariyuceJPDC,Shi11,Pande11}.

Let $G = (V,E)$ be a connected graph. Let $\sigma_{st}$ be the number of
shortest paths from a source $s \in V$ to a target $t \in V$, and
$\sigma_{st}(v)$ be the number of such $s$-$t$ paths passing through a
node $v \in V$, $v \neq s,t$. Let $\delta_{st}(v) =
\frac{\sigma_{st}(v)}{\sigma_{st}}$, the fraction of the shortest
$s$-$t$ paths passing through $v$ among all shortest $s$-$t$ paths.
The betweenness centrality of $v$ is defined by

\begin{equation}
\BC{v} = \sum_{s \neq v \neq t \in V} \delta_{st}(v).
\label{eq:bcfirst}
\end{equation}

Brandes proposed an algorithm to compute $\BC{v}$ for all $v \in V$ that is
based on the accumulation of pair dependencies over target
nodes~\cite{brandes2001}, which has $O(mn)$ complexity. Scaling the betweenness
centrality computation to large networks is impractical. To alleviate this
problem, approximation schemes~\cite{Bader07, Riondato14, Riondato16} and
parallel algorithms~\cite{Madduri2009, cohen2015all, SariyuceJPDC} are developed.

Several works have been proposed to estimate such centrality measures
using sublinear memory in the size of the graph~\cite{BoldiV13a, Priest17}.
Priest and Cybenko~\cite{Priest17} propose an algorithm that makes use of
CountSketch~\cite{charikar2002finding} algorithm. HyperBall~
\cite{BoldiV13a} work utilizes HyperLogLog variants to approximate
centralities that depends on the nodes within a certain {\it ball}
radius $r$,
such as Lin's centrality, Harmonic centrality and Closeness centrality.

However, all of those approaches focus on the network topology
information to infer the centrality, i.e., communications between each pair of
nodes are assumed to be same and non-repetitive, and need the global graph
information to highlight the most central nodes.

In our work, we propose local algorithms that can find the most critical nodes.
Our focus is on active networks where there can be non-uniform and repetitive communications between nodes, which is not studied before to the best of our knowledge.

\section{Experimental Setup}
\label{sec:exp_setup}

\subsection{Graph Instances}

\begin{table}[htb]
\caption{Graphs used in the experiments}
\label{tab:prop}
\vspace*{-1.7em}
\begin{center}
 \begin{tabular}{||c | c | c | c | c | c||}
 \hline
 Graph & \#Nodes & \#Edges & \multicolumn{3}{|c||}{Degree}\\ \cline{4-6}
  & & & min & avg & max \\
 \hline \hline
World Airports & 2,939 & 30,501 & 1 & 10.37 & 237\\
 \hline
AS-733 & 6,474 & 12,572 & 1 & 3.88 & 1,458\\
 \hline
Oregon-01 & 10,670 & 22,002 & 0 & 4.12 & 2,312\\
 \hline
\end{tabular}
\end{center}
\end{table}

We  will present results on three networks  that represent
infrastructure topologies.
Properties of these three
networks:
Airport~\cite{konect:opsahl2010b}, AS-733~\cite{snapnets} and
Oregon-01~\cite{snapnets} are presented in Table~\ref{tab:prop}. AS-733
and Oregon-01
are the graphs of autonomous systems (AS), inferred from Oregon route-
views on 1/2/2000 and 3/31/2001, respectively. In these networks, nodes
correspond to routers and the edges correspond to connections between these
routers. In our model,  ABC of a router will be the number of distinct IP
pairs that communicate through that router. Thus this will be the number of
communicating pairs that will be affected if the router were to fail.
Alternatively, one can look at the number of  distinct IPs (source or
destination), using the same algorithmic framework, but communicating pairs
provide a better handle  on the topological changes as we will describe
later.

In the Airports graph, nodes correspond to airports and there is an edge between
two nodes if there is a flight between the two corresponding airports. In this
case, our metric corresponds to how many departure/destination  pairs  go
through a given airport.

\subsection{Modeling Network Activity}
Modeling the activities on a network is a complicated problem and a challenge
in itself. For this paper, our goal is to generate instances to empirically
test our proposed methods.  Below, we first describe our data sets and then how
we generated the activity on these networks.

\subsubsection{Communication Pathways}

We assume communication between a pair of nodes follows a shortest path and the
underlying graph is connected. In our experiments, we pick a random shortest
path among all shortest paths  for each source and destination pair as their
\emph{communication path}, store this information,  and use the same path
throughout the experiment.  In practice,  communication between a pair of nodes
does not necessarily follow the shortest path, but a short path. Airlines
design their flights as such for efficiency and fewer hubs usually mean cheaper
flights, and  people prefer to take shorter flights. In communication networks,
each router in a network passes the incoming packets to a specific neighbor
router according to its routing table, where it stores the next-hop neighbor
information (the neighbor it should pass the packet to) based on the prefix of
the destination address. These tables are usually set for better communication
quality, which aligns well with shorter paths.  Unless there is a change in the
graph structure; such as, a disconnection from or a new node connection to the
network, the communication paths are unaffected. When there is a change in the
graph structure, the shortest paths for the affected nodes are recomputed.

We want to note that main contributions of this paper are independent of the
particular routing method used in the underlying system. Our shortest path
approximation can be replaced with any other path finding algorithm.  Such
changes  may affect the specific  numbers per network and nodes, but the
performances of the algorithms will be invariant.

\subsubsection{Modeling the Node Activities}

How often does a pair of nodes communicate?  We need a model that answers this
question for our experiments. In our experiments, we used a model  that assigns
a send frequency level  and a receive frequency level to each node in the
graph, and the probability of a communication from node $u$ to node $v$  is
proportional to the product of send frequency level of node $u$  to the receive
frequency level of node $v$. This is inspired by the Chung-Lu
model~\cite{ChLu02}  and  its generalization for directed graph
topologies~\cite{MeNePo06}, where  vertices are assigned in and out degrees
(analogous to active frequencies in our case), and the probability of an edge
is proportional to the product of the in-degree of one vertex and the out-
degree of the other vertex.

We have tested our approach with uniform random distribution, Gaussian
(normal), and power-law distribution. Our experiments showed that although the
actual node rankings depend on the selected distribution, the algorithm and
how it works is not dependent on it. In the rest of this paper, we present the
results using the Gaussian distribution since it is widely used to represent
and model real-life distributions in many domains.

\subsubsection{Time Window of Activities}

The measures proposed in this paper are time dependent.  The number of pairs
that communicate through a node will increase in time, but converges to the
maximum number of pairs that  communicate  through it. In our
experiments, we have used a sufficiently long time frame to observe the numbers
of convergence to avoid additional variances due to our node activity levels.
For practical purposes, we define \emph{interval} as the number of
communications over the graph.

\section{Active Betweenness Cardinality}
\label{sec:acc}

We will start this section with introducing our new proposed metric
to measure node importance that takes the network activity levels into account and does not require full access to the graph.
Next, we show how this quantity can be estimated efficiently using the Hyperloglog algorithm on a given node.
Given the ability to accurately estimate this quantity on a given node,
the next step is to find those nodes for which this quantity is the highest.
Note that we can find such nodes  by applying the same estimation technique to all nodes of the graph.
However,  that assumes we know  and have direct access to all the nodes in the graph,
which is one of the two basic premises of this work.
Moreover, we are interested in analyzing massive, distributed systems like the
Internet. As such, running a process on each router in such a system is
different than making a pass over all nodes of a graph that resides in memory.

\subsection{A New Metric for Node Criticality on Active Networks}
\label{sub:a_measure_of_activity_importance}

In this section we  introduce {\em Active Betweenness Cardinality} (ABC) as a new metric
that measures node  criticality not only based on topology,  but  on the activity level of the network.
The proposed method also has algorithmic advantage that it can be computed based on
only local information and  we do not need to know the exact topology of the network.

Finding critical nodes in a graph has been studied in depth in the literature, with centrality based measures being a major thrust.
Our proposed metric,  inspired by the centrality ideas, betweenness centrality
in particular, but it differs  from these approaches in two major ways.
Present methods for centrality rely on two assumptions:  (1) centrality of a node depends only on
the topology of a graph and do not take into account the dynamics on the graph
or how the graph is being utilized, and  (2) we  know the exact topology of the
whole graph.

These assumptions restrict the effectiveness and applicability of these methods.
First, we cannot assume there will be an interaction between all
pair of nodes on a large graph at the same level or frequency. This may be due
to lack of interest (e.g., not every user visits every page on the Web) or some
nodes provide identical services and users only interact with the nearest such
server (one goes to a close-by hospital, not necessarily all the hospitals).
One question we could ask is can we develop approaches with models that
comprise probabilities of  interactions between every pair of vertices? While
this sounds good at first, building such a vast model with high confidence can
be as burdensome as  computing the centralities themselves. It should be noted
that  centrality measures were initiated by the social science studies on much
smaller networks, where neither of the two limitations above apply. But one
should be careful before applying the same ideas to current datasets, where the
graphs are much larger  and  comprise nodes with identical functionalities.

The second problem is the assumption that we know the exact topology of the
network. Many networks such as the Internet are vast, distributed, and
autonomous. Moreover, both their topology and the dynamics on them keep
evolving. Determining their exact topology or even predicting their topological
features is far from trivial.

 Here we propose a new metric,  active  betweenness cardinality, that avoids
 both  problems.

\begin{definition}
{\em Active Betweenness Cardinality} (ABC) of node $v$ for time, period
$[t_s, t_f]$,
\mbox{$ABC(v, t_s, t_f)$}  is the number of  distinct $\langle s, d, f \rangle$ tuples for
a go-through
node $v$, where $s$ and $d$ are the source and destination nodes of the
communication, which we will also call as  {\em transaction}, and $f$
represents the features associated with the transaction.
\end{definition}

Note that, this metric is time dependent. However, as the difference between
the start and end times increase, the metric converges.
In our experiments, we have used   the estimate after conversion,
and will not specify the time intervals in the remainder of the paper.
The features $f$, can be used to focus the analysis on certain types of
activities in the network.
For instance, we can only investigate trucks,  not all vehicles,
or we can focus only on http traffic in a network.
In our experiments we have used any specific features.
We will only refer to the ABC value of a vertex, when the time interval is
clear from the context.

\subsection{ Computing ABC Accurately and Efficiently }
\label{ss:hll}

In this section, we will show how the ABC metric can be computed efficiently,
using only local information. Computing the ABC of a node corresponds to the
count-distinct problem, which we have presented earlier in
Section~\ref{sec:HLL}. We  are interested in handling massive graphs with
heavy traffic, which prohibits exact calculations due to speed of data
arrival and  memory requirements. Instead, we will use provably accurate
approximation algorithms, specifically the HyperLogLog (HLL) algorithm, which
was also discussed in Section~\ref{sec:HLL}.  We will start with verifying
that this techniques provides accurate estimations, as this technique will be
the building block for other techniques in the remainder of the paper. Note
that this technique does not require  any information about  the remainder of
the network. All we need to know is the source and destination of each
transaction that go through the node.

\begin{figure}[h]
\centering
  \includegraphics[width=0.48\textwidth]{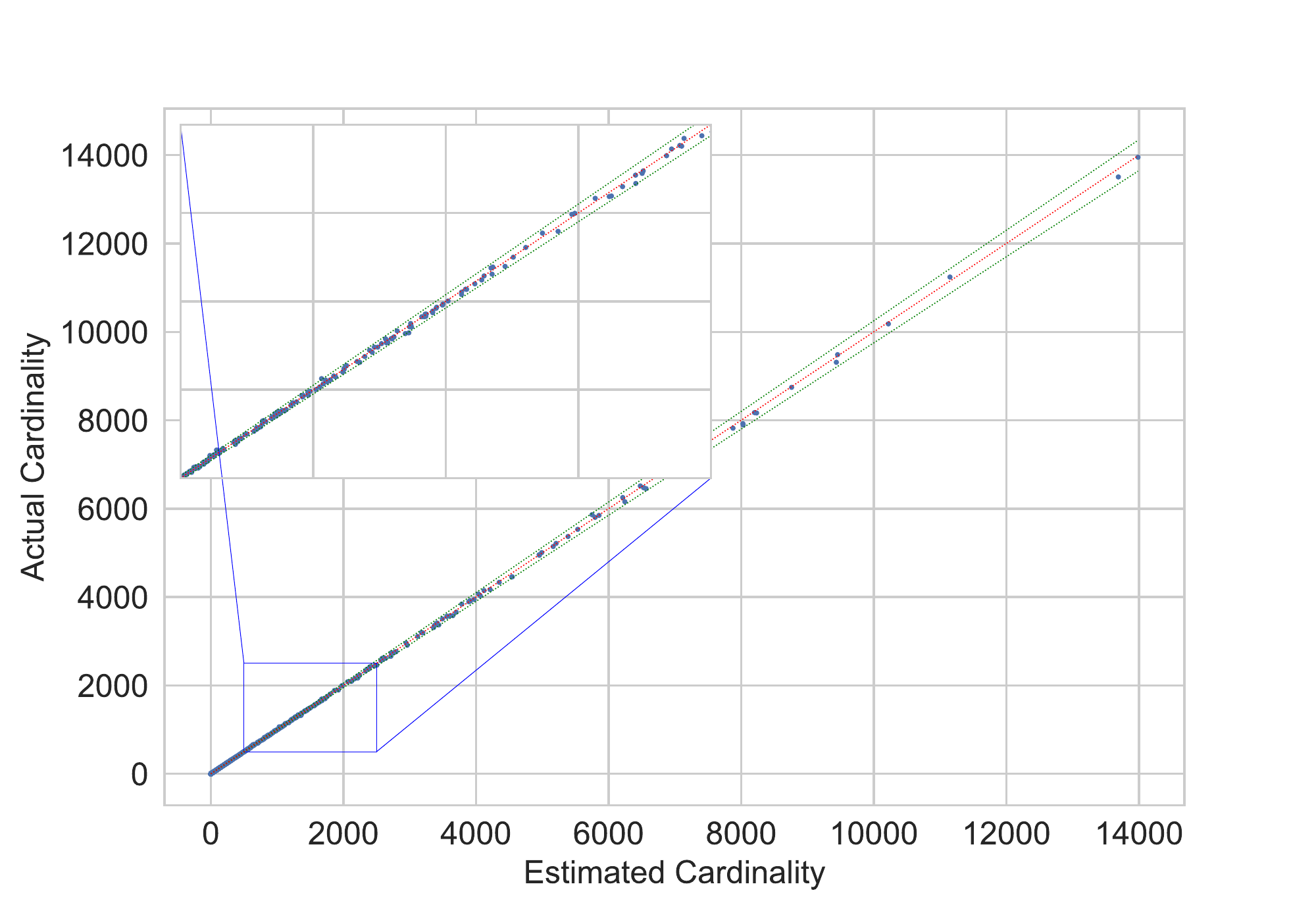}
  \vspace*{-3ex}
  \caption{The actual unique number of (src,dest) pairs for each node
  vs HyperLogLog estimates. The red line is $y=x$ and green ones are $\pm2\%$ error ranges.}
  \label{fig:accuracy}
\vspace*{-1ex}
\end{figure}

The results of our experiments are presented in Figure~\ref{fig:accuracy}.
In this figure each  data point corresponds to the ABC of a different vertex
in the World Airports Graph for one interval of size 50,000 transactions. The
x-axis  corresponds to the exact cardinality  computed using the set
approach, while the y-axis corresponds the approximation provided by the HLL
algorithms.  The red line corresponds to the ideal solution such that $x=y$,
and the two green lines  correspond to 2\% over and under approximation,
i.e., y=1.02x and y=0.98x.

\vspace*{-2ex}

As can  be seen in Figure~\ref{fig:accuracy}, the hyperloglog estimator
consistently provides good estimates. always within 2\%. We have run these
experiments 20 times with different random seeds to generate different
instances. Average Pearson Correlation Coefficient for \emph{Actual} and
\emph{HLL estimated} cardinalities was 0.99994.  The accuracy of HLL
estimates is well-studied, thus we  will not be presenting any further
experiments here. But we hope that this  experiment helps those who are not
familiar with HLL techniques and show that the HLL algorithms works well in
this case as well.

\subsection{Finding Nodes with Highest ABC Scores}
\label{sec:finding_nodes}

The results of the previous section show that we can efficiently estimate
the ABC value of a given
node.  But how do we  find the nodes with highest scores when we do not even
know about the presence of nodes in the rest of the graph. If we knew  and
had access to all the nodes in the graph, we could have computed the
estimates on all nodes to identify the most critical nodes (those with
highest ABC scores).  However, this is not feasible in a vast  and autonomous
system like the Internet. Here, we propose a set of heuristic algorithms that
searches for the nodes with highest ABC scores, in a distributed
fashion, starting from a handful of nodes.  Below, we will describe
our search strategies and experimentally show their effectiveness.

\subsubsection{Search Strategies}

Our techniques are based on choosing one of the neighbors of the current node
that is most promising to have a higher ABC score than the current node and
repeat. In all cases, we avoid backtracking to the node, where we came from.
Starting from $C$ seed nodes, all of the search strategies iteratively
selects next $C$ nodes to continue to search. We call each iteration a {\em
hop}.

We show a number of approaches to traverse the graph and find important
nodes: (i)~traverse toward the neighbor with highest cardinality (via the
edge ABC value), (ii)~biased random walk to one of the neighbors where random
walk probability of each neighbor is proportional to the cardinality of the
edge between these two nodes, (iii)~coordinating among  multiple searches to
investigate best overall alternatives.

{\em Jump to Best Neighbor (BN)}:
For each neighbor of the current node, or in another words, for each edge of
the current node, we maintain an HLL estimator that counts the distinct pairs
that communicated through that neighbor/edge.
Intuitively, if we have an important neighbor, it would acquire most of our
communication output. Alternatively, the current node is the important
neighbor of that node that provides most of its input.
Following that intuition, we select the
out-neighbor that has the highest cardinality on the edge that connects the
current node to that node as the candidate.
We also propose a variation of BN that excludes selecting the nodes with
only one out-neighbor, and we call that BN-1x.

{\em Jump with Random Walk (RW)}:
Jumping to the highest neighbor could get us stuck at a local extrema where
we do not have anywhere else to jump since we do not allow jumping back to a
previously observed node. For this purpose, we implemented a Random Walk-like
approach where the neighbors have probabilities proportional to their
corresponding HLL estimators on the edges. For example, let us consider a
node A with out-neighbors \textbf{X}, \textbf{Y}, \textbf{W}, and
\textbf{Z}. The HLL
estimations on the edges are \{60, 32, 80, 28\} respectively. With
\textit{BN} strategy, we would jump to \textbf{W} and we might miss the
chance to see \textbf{X}. We normalize these values and get
\{0.3, 0.4, 0.16, 0.14\}. Then, we randomly pick one of the neighbors with
this probability distribution as the candidate. Similar to BN, propose a
variation of RW that excludes selecting the nodes with only one out-neighbor,
and we call that RW-1x.

{\em Biggest Collective Neighbor Selection (BCN)}:
An alternative approach is instead of each node selecting one node to
jump-to, collectively they can decide which nodes should be next subset of
nodes to evaluate. This will speed up the convergence in cases where one of
the subset nodes have all important neighbors but all the others are not
necessarily as important. For example, consider a scenario that the set of
nodes we are evaluating are \{\textbf{A}, \textbf{B}, \textbf{C},
\textbf{D} \}, and all the neighbors of node \textbf{A} has
cardinalities in the scale of ten thousands
and the neighbors of \textbf{B}, \textbf{C}, and \textbf{D}, have
cardinalities in the scale of thousands.
In such a case, it would make much more sense to select the nodes of new
subset all from the neighbors of node \textbf{A} instead of picking one
neighbor from each node. Converting this to a more general form, we select
the top $C$ non-observed neighbors of all neighbors of current subset.

\subsubsection{Quality Metric for Search Strategies}

In this section we define the quality metrics for the presented strategies to
find critical nodes.
First metric is one-to-one correspondence of the critical nodes found and
the \emph{ground truth}. Second, the total (sum of)
cardinality value found compared to the \emph{ground truth}.
Since what constitutes a ground truth depends on the active communication and
activity of
nodes, we need to decide what is ground truth for that specific configuration
and settings. Hence, we generate a baseline from 400 runs with the same flow
probability distributions, then average the ABC scores of each node to get
a baseline for our experiments.

{\em Top-$K$ Found Matches}:  For all algorithms, we compare the top $K$
nodes found with the top $K$ in the baseline. If the node found by the
algorithm is
among the top $K$ in the baseline, then we count this as a hit (success). So,
perfect result would be to find $K$ out of $K$. We have run the experiment
sets for 20 times and averaged the results.

{\em Total Cardinality of Top-$K$ Found}: If the cardinalities of the most
critical nodes are not distinguishably different, comparing top-$K$ vertices
against ground truth will not be fair. For such cases we simply compare the
the sum of the total cardinality of top-$K$ found against the sum of the
total cardinality of top-$K$ in baseline.

\subsubsection{Experimental Results}

\begin{figure*}[ht]
\centering
  \subfigure[AS Graph (Top to bottom: RW, RW-1x, BN, BN-1x, BCN)]{
    \includegraphics[width=0.8\textwidth]{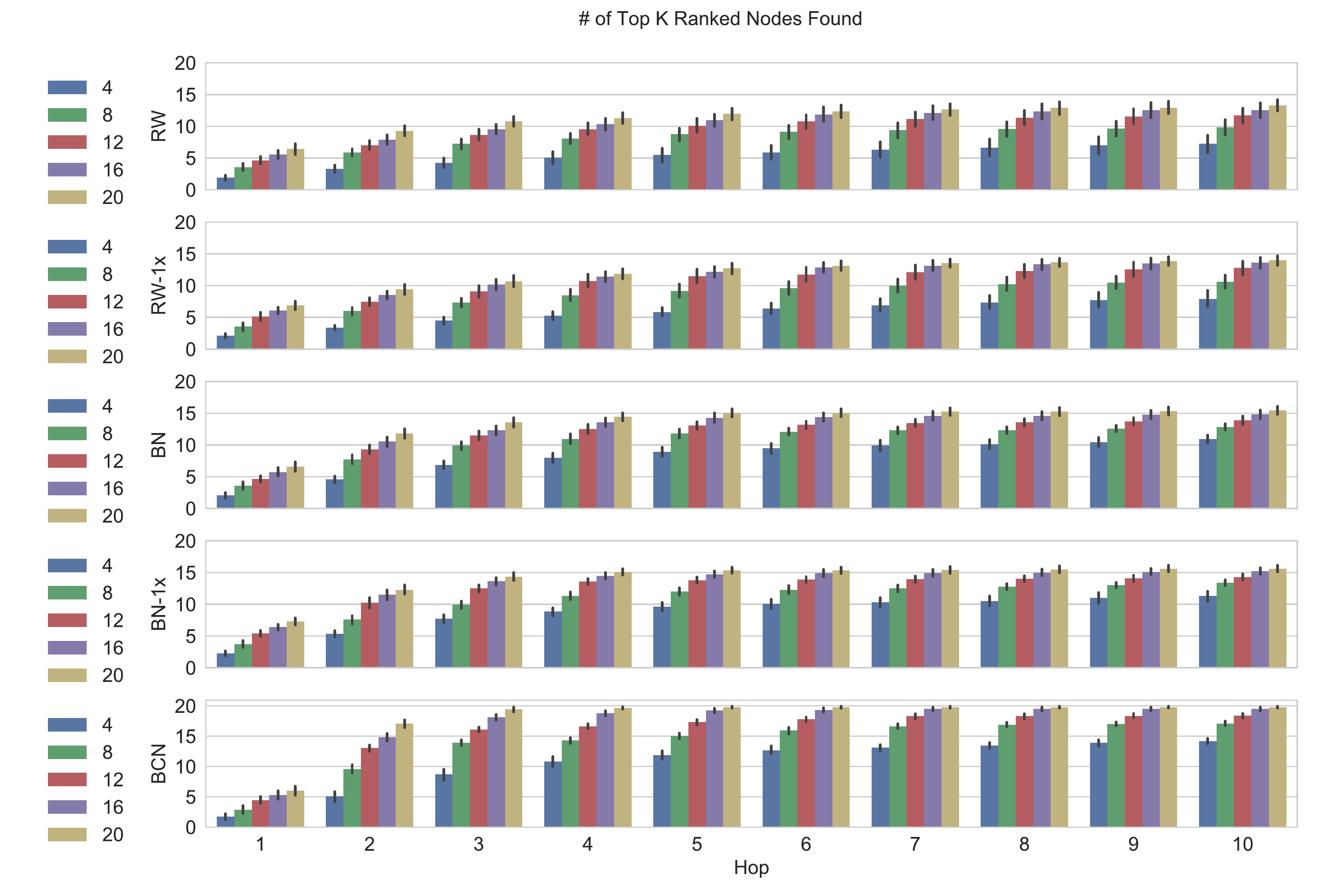}
    \label{fig:topKas}
  }
  \subfigure[World Airports Graph (BCN)]{
    \includegraphics[width=0.8\textwidth]{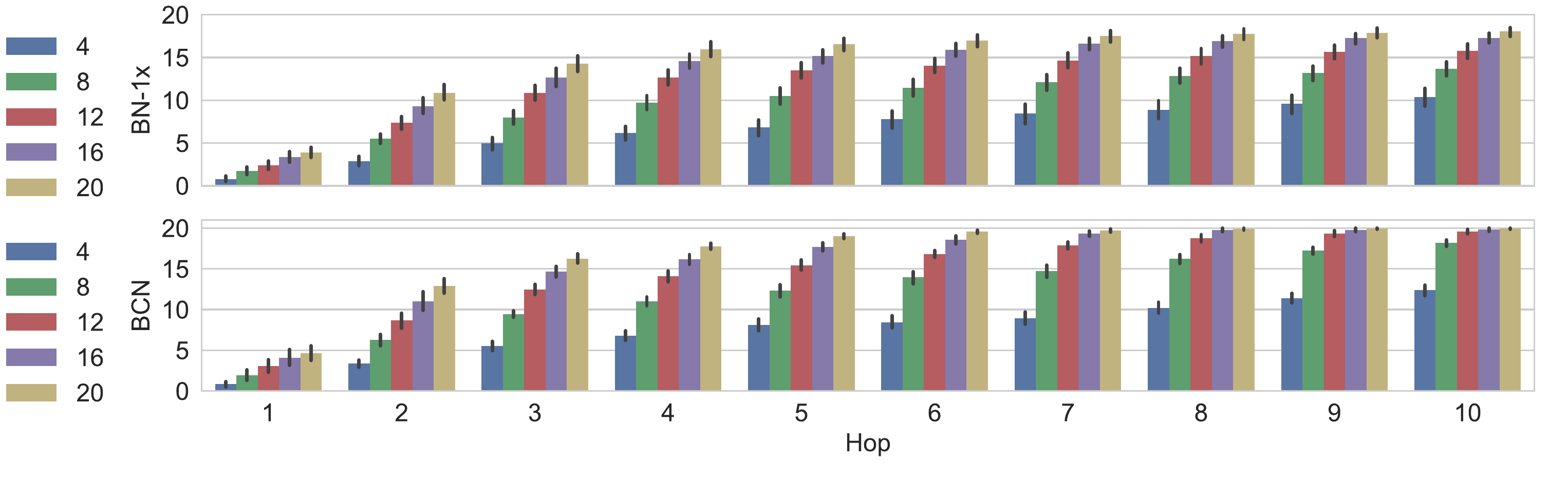}
    \label{fig:topKairports}
  }
  \subfigure[Oregon Graph (BCN)]{
    \includegraphics[width=0.8\textwidth]{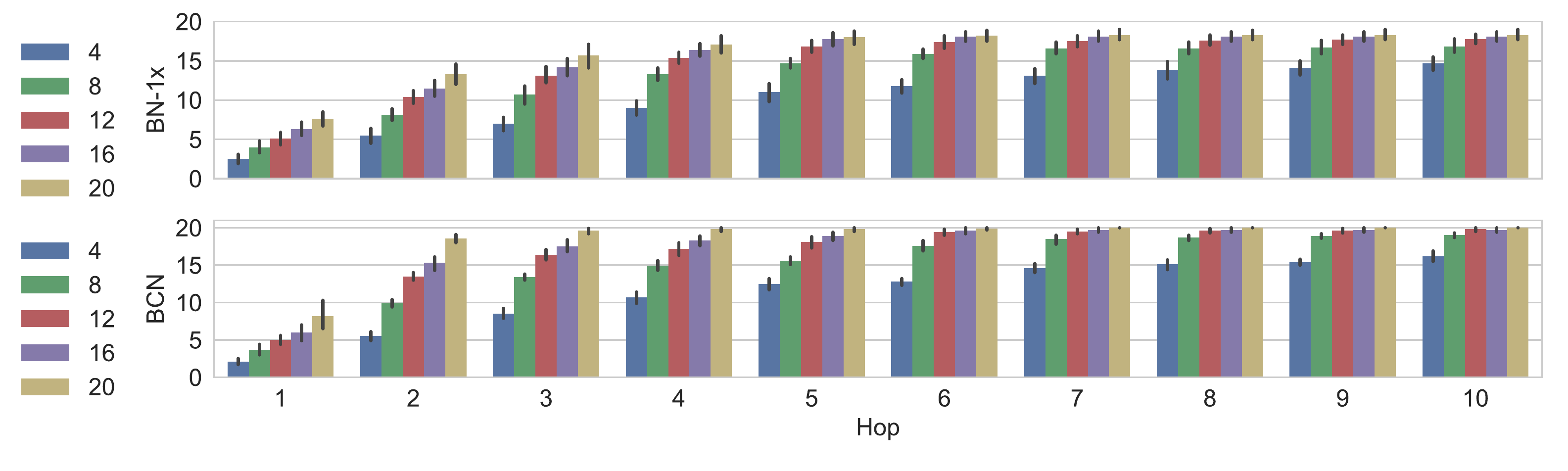}
    \label{fig:topKoregon}
  }
  \vspace*{-1.7em}
\caption{Top-$K$ found matches for five search strategies.}
\label{fig:topK}
\end{figure*}

Figure \ref{fig:topK} shows the comparison of the search strategies for the
three test graphs we have. The first two strategies are the random walk
selections RW and RW-1x, second two are the best neighbor selections, BN and
BN-1x, and the last one is
the collective best neighbor selection strategy, BCN. Figure~\ref{fig:topKas}
shows all five alternatives for AS Graph. Figure~\ref{fig:topKairports} and
\ref{fig:topKoregon} presents only the BN-1x and BCN to show the trend of the
algorithms do not change much when different graphs are in question. Each
bar represents the number of nodes in the search set, i.e., $C =
\{4,8,12,16,20\}$.

As seen in Figure~\ref{fig:topK}, as expected, as the search set size,
$C$, gets bigger, so the quality of the result. However, even with modest
sizes of $C$, where $C \approxeq K$, we can identify a large portion of the
critical nodes. In addition, our search strategies requires small number of
hops to stabilize, and they succeed to identify large portion of the critical
nodes with RW and BN variants, and all top-$K$ with BCN. The figure also
depicts that the quality of the approaches increase from top to bottom, having
BCN the best and BN-1x the second best while random walk approaches have
slightly worse results. Moreover, it shows that the results tend to converge in
less than 7 hops for all 3 of the graphs. Finally, BCN also
converges much quickly than the other variants.
One could argue that the BCN strategy requires a collective selection, thus,
more difficult deploy  as a distributed strategy. We show the results of BN-1x together
with BCN for all 3 graphs for this strategy to show one can use BN-1x instead
of BCN and still have similarly high-quality result.

\begin{figure*}[ht]
\centering
  \subfigure[AS Graph (Top to bottom: RW, RW-1x, BN, BN-1x, BCN)]{
    \includegraphics[width=0.8\textwidth]{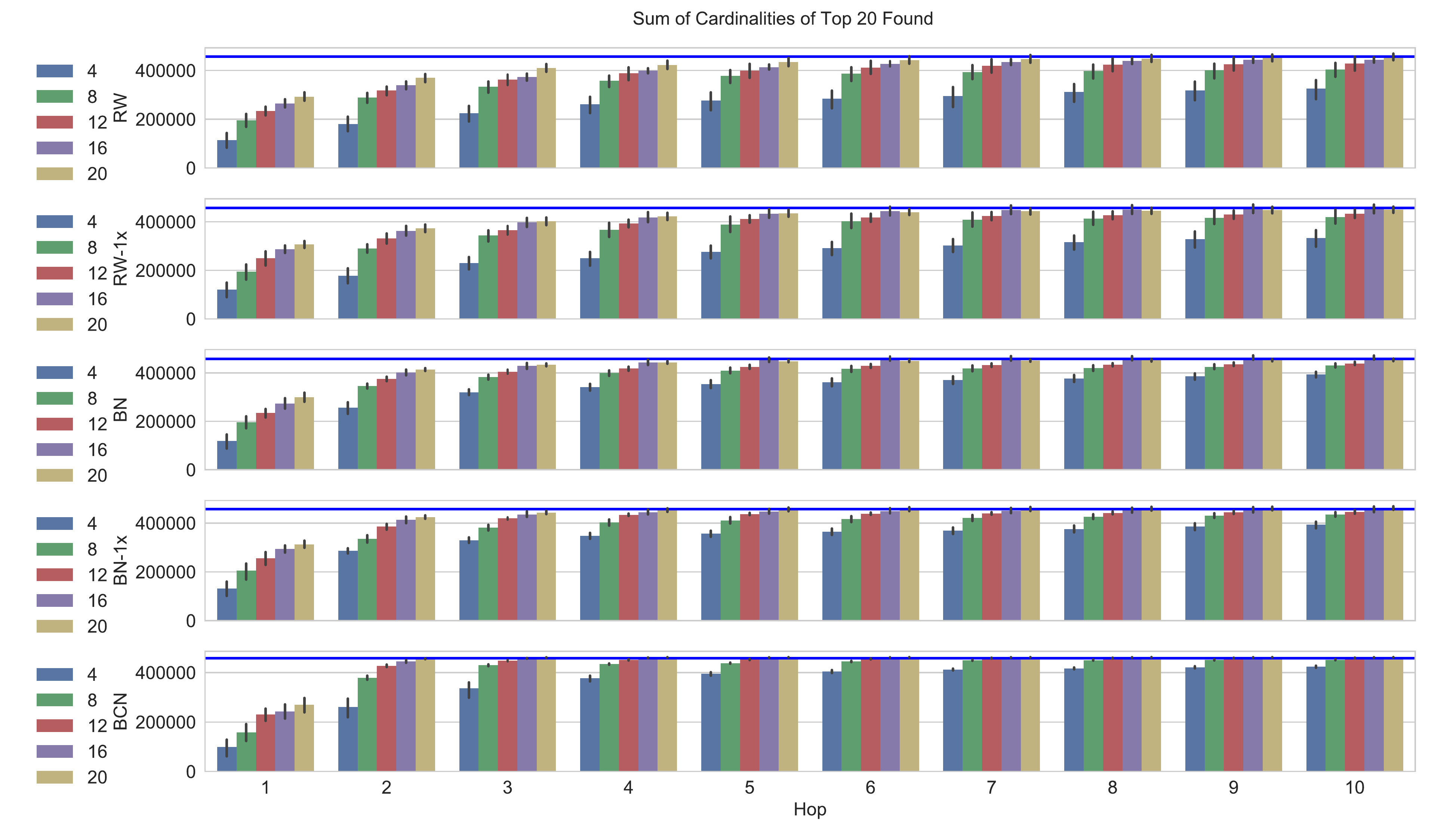}
    \label{fig:cardSumas}
  }
  \subfigure[World Airports Graph (BN-1x, BCN)]{
    \includegraphics[width=0.8\textwidth]{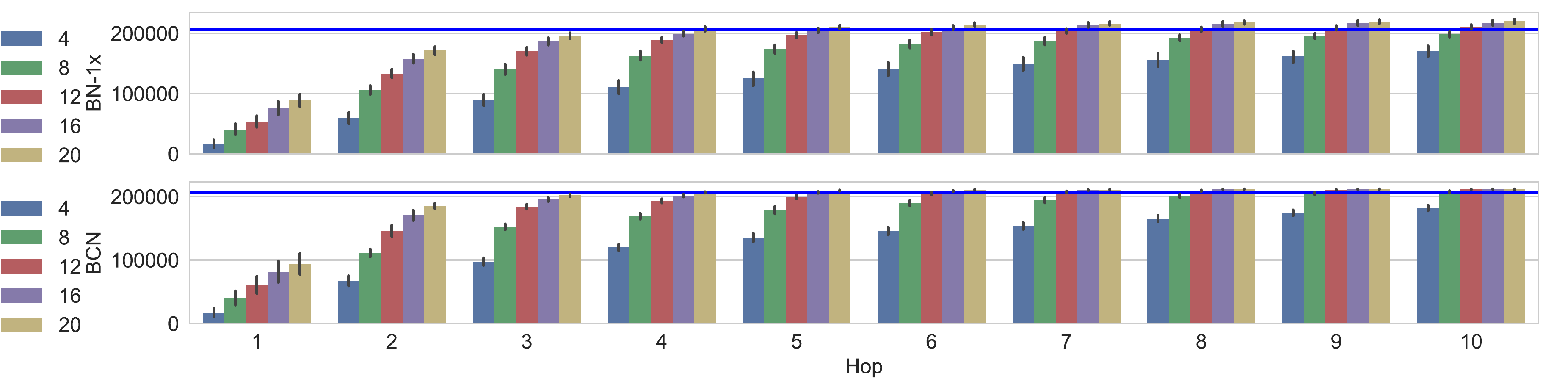}
    \label{fig:cardSumairports}
  }

  \subfigure[Oregon Graph (BN-1x, BCN)]{
    \includegraphics[width=0.8\textwidth]{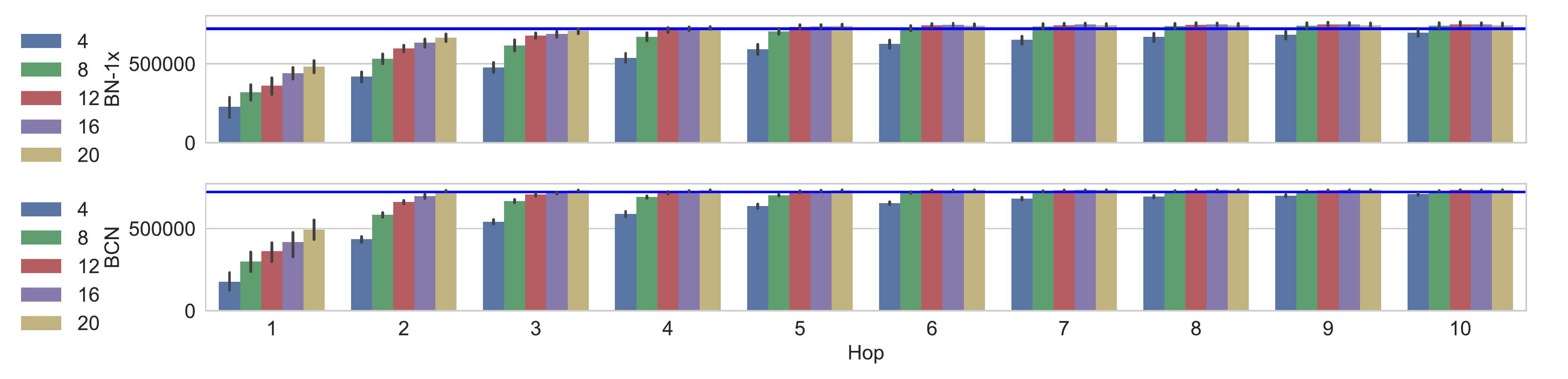}
    \label{fig:cardSumoregon}
  }
\vspace*{-1.7em}
\caption{Total Cardinality of Top-$K$ Found for five search strategies.}
\label{fig:cardSum}
\end{figure*}

Figure~\ref{fig:cardSum} shows a comparison of the search strategies using
our second metric, total cardinality of top-$K$ found, which is
displayed as blue solid line on the chart. As in Figure~\ref{fig:topK}, we
displayed all five search strategies for AS graph, and strategies BN-1x and BCN
for the other two graphs.
The total cardinality
found by each algorithm gets very close to the one identified in baseline,
telling us top critical nodes have very similar cardinalities, thus those the
BN and RW algorithms missed in Figure~\ref{fig:topK} from top-$K$ may not be
much more critical than what those nodes found in terms of the nodes that
depend on them.
Similar to previous experiment, if fully distributed algorithm is preferred,
one can use BN-1x instead of BCN and still have high-quality result.

\section{Detecting Changes in Network }
\label{sec:detecting_changes}

In this section, we showcase how the ABC values can be used to monitor a network
and detect significant changes in the network topology.  We first show how to detect the
failure of a critical node by using a small number of  sensors, then we discuss how to locate those nodes.
We restrict ourselves to detecting failures of nodes with high ABC scores (e.g., top 200 or 400),
as failures of  such nodes make an overall impact on the functionality of the network.
The question is if this change can be captured  by our metrics by using a limited number of sensors.
Failure of  nodes with low ABC scores on the other hand is much harder to detect, as their absence is not expected
to make an overall impact, and require sensors that are very close.

\begin{figure}
\centering
\includegraphics[width = 0.48\textwidth]{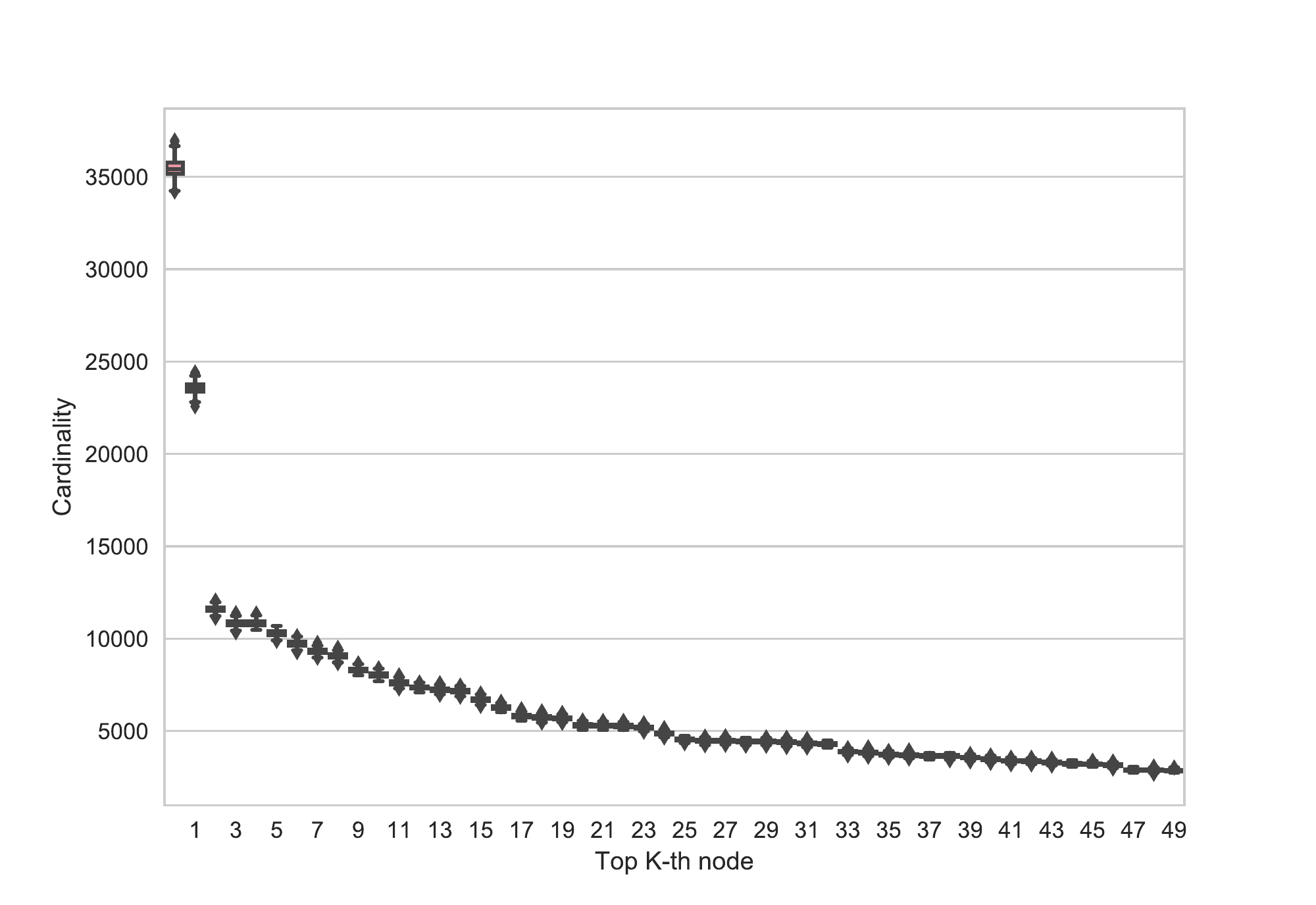}
\vspace*{-1.7em}
\caption{ The ranges of Top 50  ABC values for the Airports graph for 200,000
transactions after 400 experiments with the same communication probability
distribution.  The box corresponds to the first and third quartiles. Whiskers
show the range for $\pm 1.5\times IQR$, where $IQR$ is the difference between
the third and first quantile values. Any outliers would have been marked with
plus signs. }
\label{fig:baseline}
\end{figure}

The critical observation for our methods is that the ABC values provide a stable baseline  with very small variance,
when the underlying network is stable.
However, the same ABC scores are sensitive to the changes in the network, as we will show later.
Here we first want to present the stable, low-variance nature of ABC scores in the network.
Figure~\ref{fig:baseline} shows the top 50 ABC values of the
Airports graph for an interval size of 200,000. Experiments are repeated 400 times with the same
communication probability distribution.  In this figure, the box corresponds to the first and third quartiles.
Whiskers show the range for $\pm 1.5\times IQR$, where $IQR$ is the difference
between the third and first quantile values.
Any outliers would have been marked with plus signs, but as can be seen  there are no outliers, and the variances
are extremely small, forming a good baseline to detect any changes.

\subsection{Detecting the Removal of Critical Nodes}
We first discuss our approach to characterize what is normal and how we predict an anomaly based on only positive samples.

\subsubsection {Characterizing Normal}
One way to detect anomalies is to generate many examples that describe the
system at its normal state and use them to build a distribution to define the
normal. In our experiments we generate 200 examples and describe the state of the system as a 20 dimensional vector, which are the ABC values of critical nodes chosen randomly out of the top  200 nodes.   We have used  these examples to train  OneClassSVM\cite{scikit-learn} to form a model, and for each test case, it checks whether the description fits into \textit{normal} or
\textit{anomalous}. We generated test cases  by removing  (10, 20, 30, or 50 nodes) out of the top 200 nodes.

Overall, we observed 80\% accuracy.
Such an accuracy is high for an approach where only positive
examples are provided.  However, test cases comprised  significant
failures that involve with many failed nodes.  We observed that affects of failures remain local, even when significant.
Failure of one node may double the ABC value of another node, which is
sufficient. However, this effect can be minimal within
the full-scale system, as it is absorbed locally.  This motivated us to use
supervised approaches which we explain next.

\subsubsection{Binary Classification with failure examples}

We train the algorithm with both the baseline and a set of results for anomaly
cases where we also provide a label for each case, i.e. $0$ for normal, and $1$
for anomaly. Anomalies were based on removal of  (10, 20, 30, or 50 nodes) out of the top 200 nodes.
We have experimented with
classification algorithms SVM (with linear and Gaussian (rbf) kernels), Linear
Regression, Decision Trees, ExtraTrees, AdaBoost, Neural Networks (L-BFGS, Adam)
and Ensemble (Weighted Voting based) of these in the python library
scikit-learn\cite{scikit-learn}. We have found SVM with Gaussian kernel to be effective and fast.

We do 10-fold cross validation~\cite{scikit-learn}. In our experiments we have seen the supervised learning have  near $100\%$ accuracy in sensing the removal of
a critical node subset. With this success, we investigated whether we can detect the location of the failure, not only the presence of the failure.

\begin{figure}
\centering
\includegraphics[width = 0.48\textwidth]{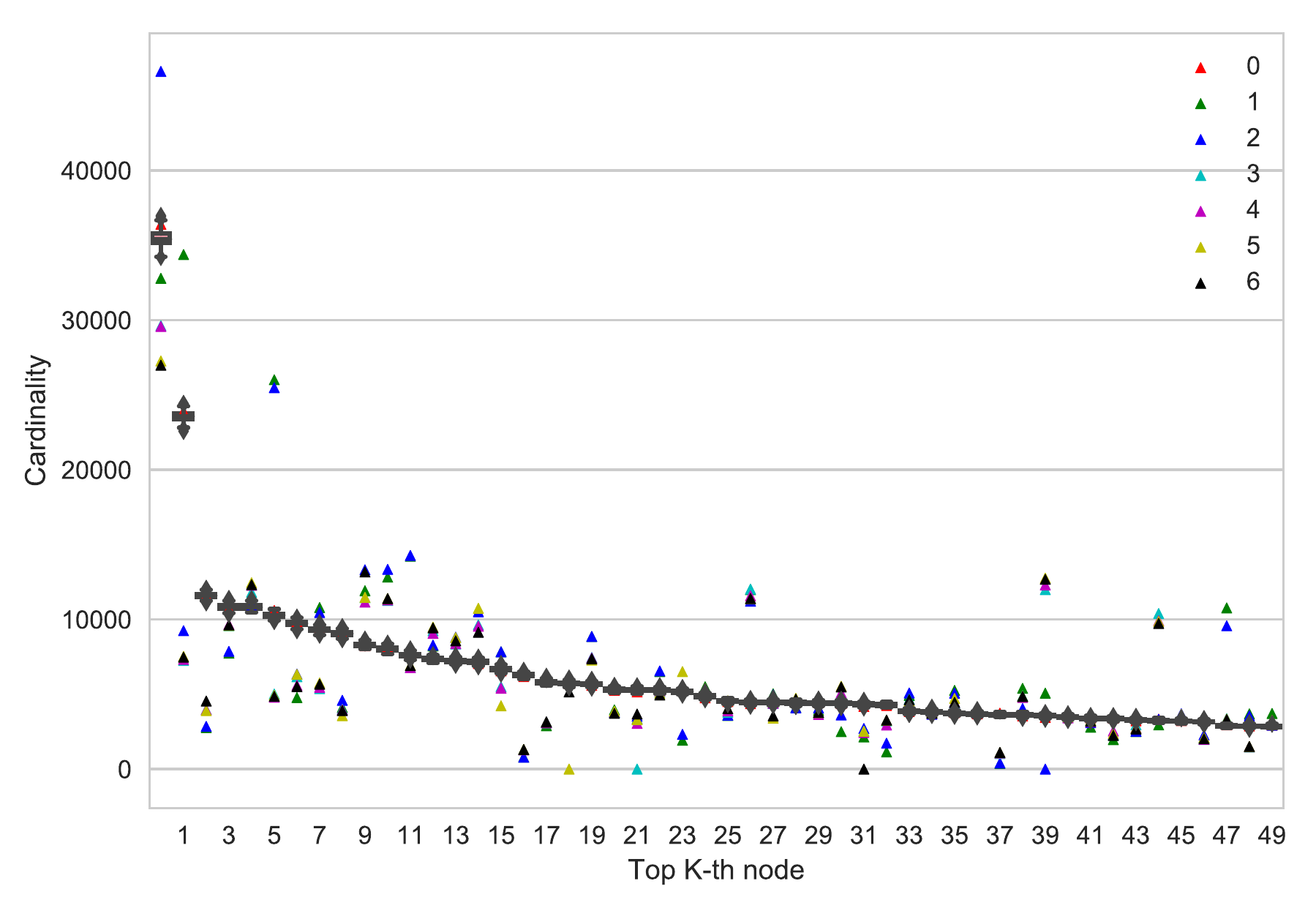}
\vspace*{-1.7em}
\caption{Box and Whiskers range for regular case and 7 instances of removal of
critical nodes. The box corresponds to the first and third quartiles. Whiskers
show the range for $\pm 1.5\times IQR$, where $IQR$ is the difference between
the third and first quantile values. }
\label{fig:baseline_plus}
\end{figure}

\subsection{Locating the Failure}
\label{sub:detecting_specific_changes_in_network}

Now that we can detect the presence of a failure, can we  go a sep further and detect  which node failed?
 The results in Figure~\ref{fig:baseline_plus} motivate our work.  In this figure, we show how  the top 50  ABC values  change after failure of the top 6 nodes.  In this figure label 0, corresponds to  no failures, and the labels 1-6  correspond to  failure of the node with the corresponding  ABC rank.  We plot these results on the baseline of Figure~\ref{fig:baseline}.  As the figure shows, each failure pushes at least one ABC value significantly out of range.  The question is whether we can identify a combination  that can help us uniquely identify each failure.

The problem would be trivial, if we have sensor on each of the top K nodes.  But we want to be able to infer beyond what we can observe directly.
So instead we use the nodes ranked from $10{th}$ to $30{th}$  to predict failures of top 7 nodes. We assume single failures and only one important node from top 7 nodes is removed at one time. We train our classification algorithm with the baseline
and a sample set of 200 instances for the removal of each of top 7 nodes.
We test our algorithm with 10-fold cross validation~\cite{scikit-learn}.

\begin{figure*}
\centering
\includegraphics[width = 0.8\textwidth]{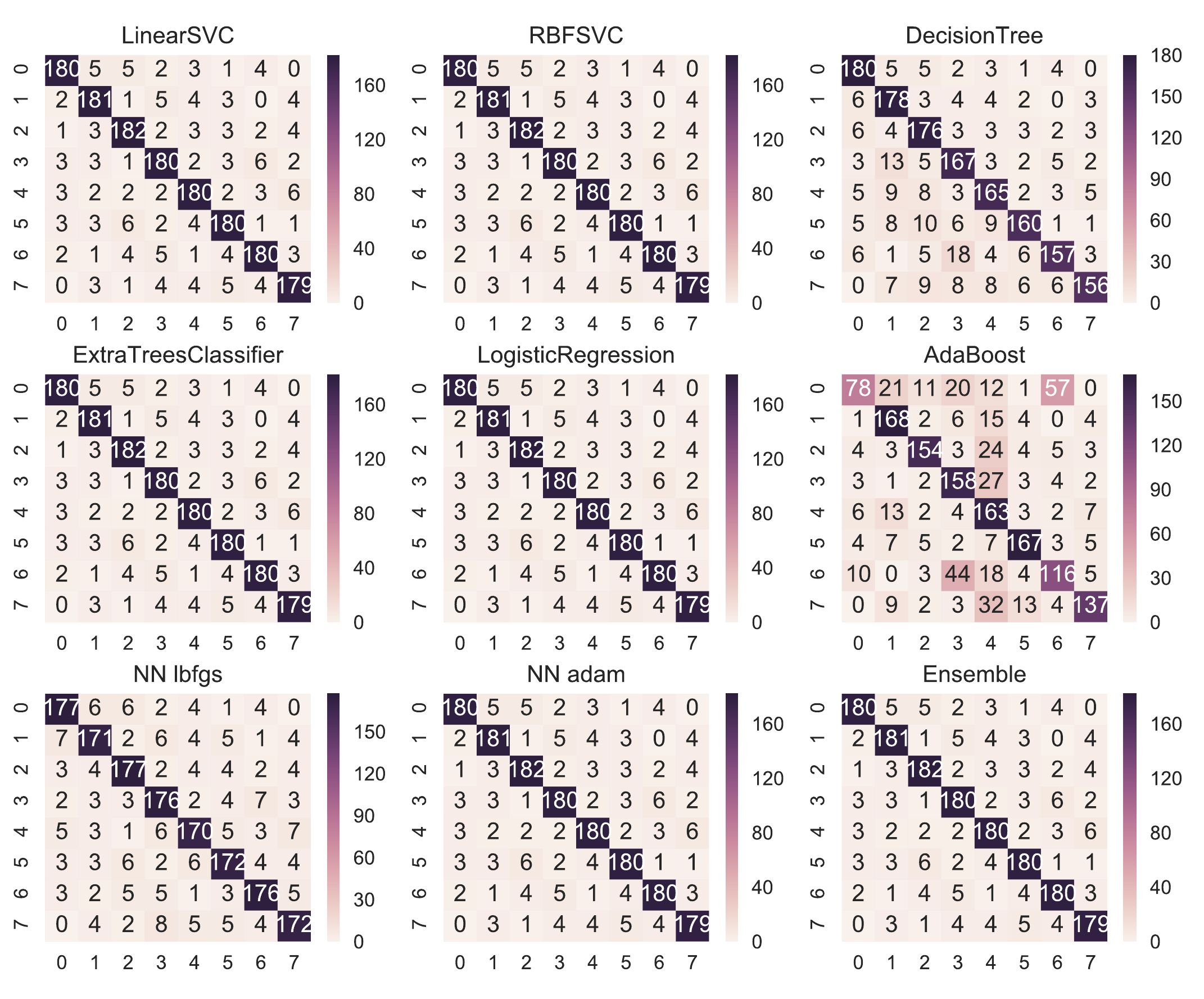}
\vspace*{-1.7em}
\caption{AS-733 Graph; Comparison of classification approaches.
Detecting which one of top K nodes is disconnected using the
top M nodes. ( k = 7, m= 40 in this case).}
\label{fig:as_All_ML_Methods}
\end{figure*}

\begin{figure*}
\centering
\includegraphics[width = 0.8\textwidth]{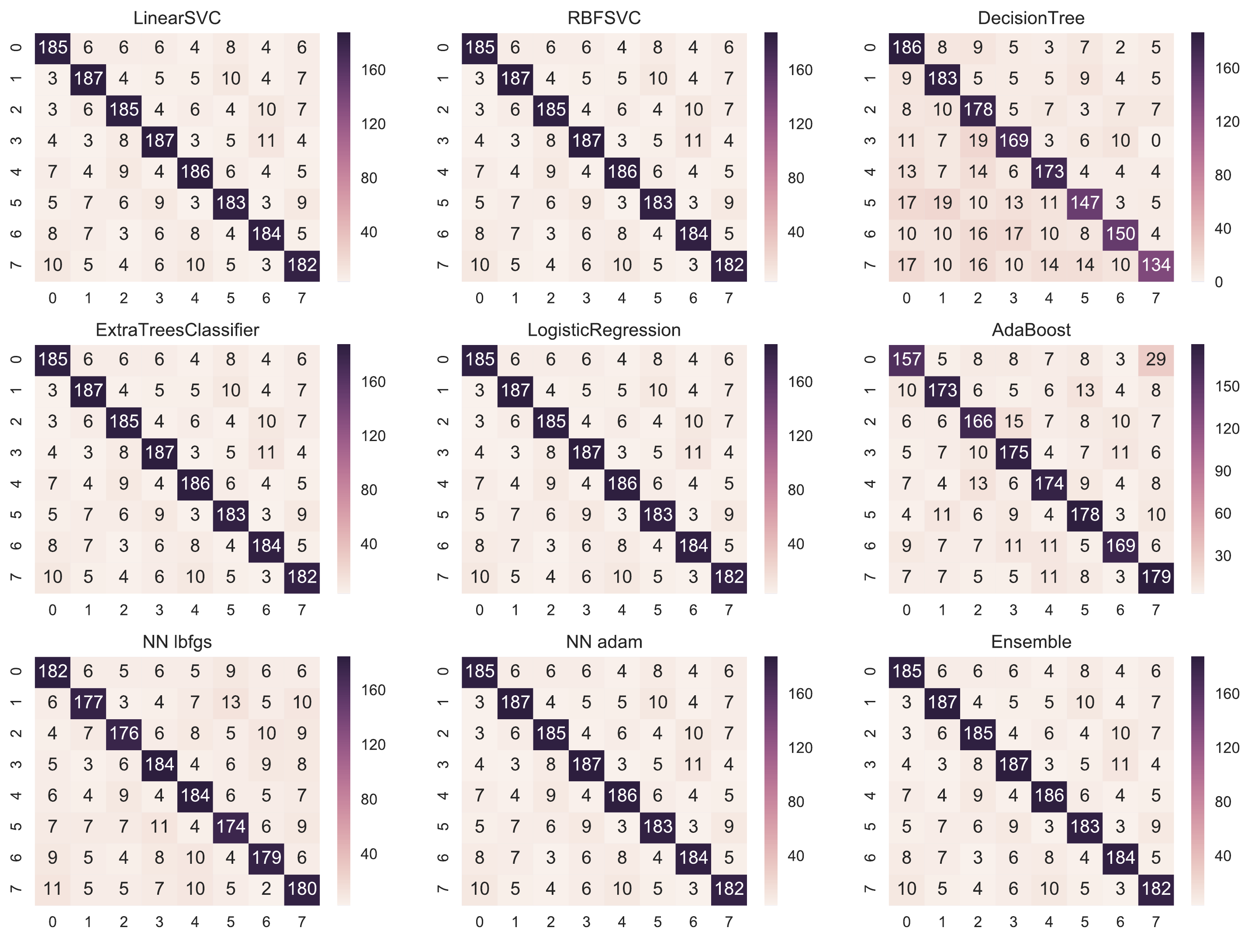}
\vspace*{-1.7em}
\caption{Airports Graph; Comparison of Machine Learning Approaches.
Detecting which one of top K nodes is disconnected using the
top M nodes. ( k = 7, m= 40 in this case).}
\label{fig:All_ML_Methods}
\end{figure*}

Figures \ref{fig:as_All_ML_Methods} and \ref{fig:All_ML_Methods} show the confusion matrices for 8 labels
(Zero is when there is no change in the graph, i = 1 through 7 is the removal
of top $i-{th}$ node) for 9 different classification algorithms we
experimented with (namely, Linear SVC (Support Vector Machines with Linear
Kernel), RBFSVC (Support Vector
Machines with RBF (radial basis function, Gaussian) Kernel), Decision Trees, Extra Trees,
Logistic Regression, AdaBoost, Artificial Neural Networks L-BFGS (Limited
Memory Broyden-Fletcher-Goldfarb-Shanno) algorithm, Artificial Neural Networks
(Adam method), and lastly the weighted Voting Ensemble).  We used the
implementations of these algorithms in
 the python library scikit-learn\cite{scikit-learn}.

These confusion matrices show the count of real and predicted label count
for each instance in the test set. Having a diagonal matrix (where all entries
outside the main diagonal are zero) is the perfect classification where the
algorithm manages to successfully predict the actual label of each test case.
Thus, Figures~\ref{fig:as_All_ML_Methods} and~\ref{fig:All_ML_Methods} show that the SVM Classifier with RBF kernel gives the best
results. (Although some other algorithms also give the same amount of correct
answers in this experiment, we have seen that RBFSVC works fast and better
overall including the other sensing/detecting experiments.)

Overall,  the results show that  the ABC values can help accurately identify  a missing node among  the top nodes. It is worth noting that a na\"ive guessing approach would have ${\sim}15\%$ accuracy with a balanced training set.

\subsubsection{Minimizing the Number of Sensors}

If we can detect changes in top  $7$ nodes using top $20$
nodes excluding top 7, then the next question to ask is, "What is the minimum
number of nodes to be observed  to detect changes in the
top $K$ nodes?" For this, we disconnect top $K$ nodes one by one and
try to detect the changes using the remaining top $M-1$ nodes' ABC values.
Then, create a matrix to present which ones can classify the changes caused by
which other nodes. Later, this matrix can be used to find the minimum required number
of nodes to be observed to detect changes on all top K nodes.

\begin{figure*}
\centering
\includegraphics[width = 0.8\textwidth]{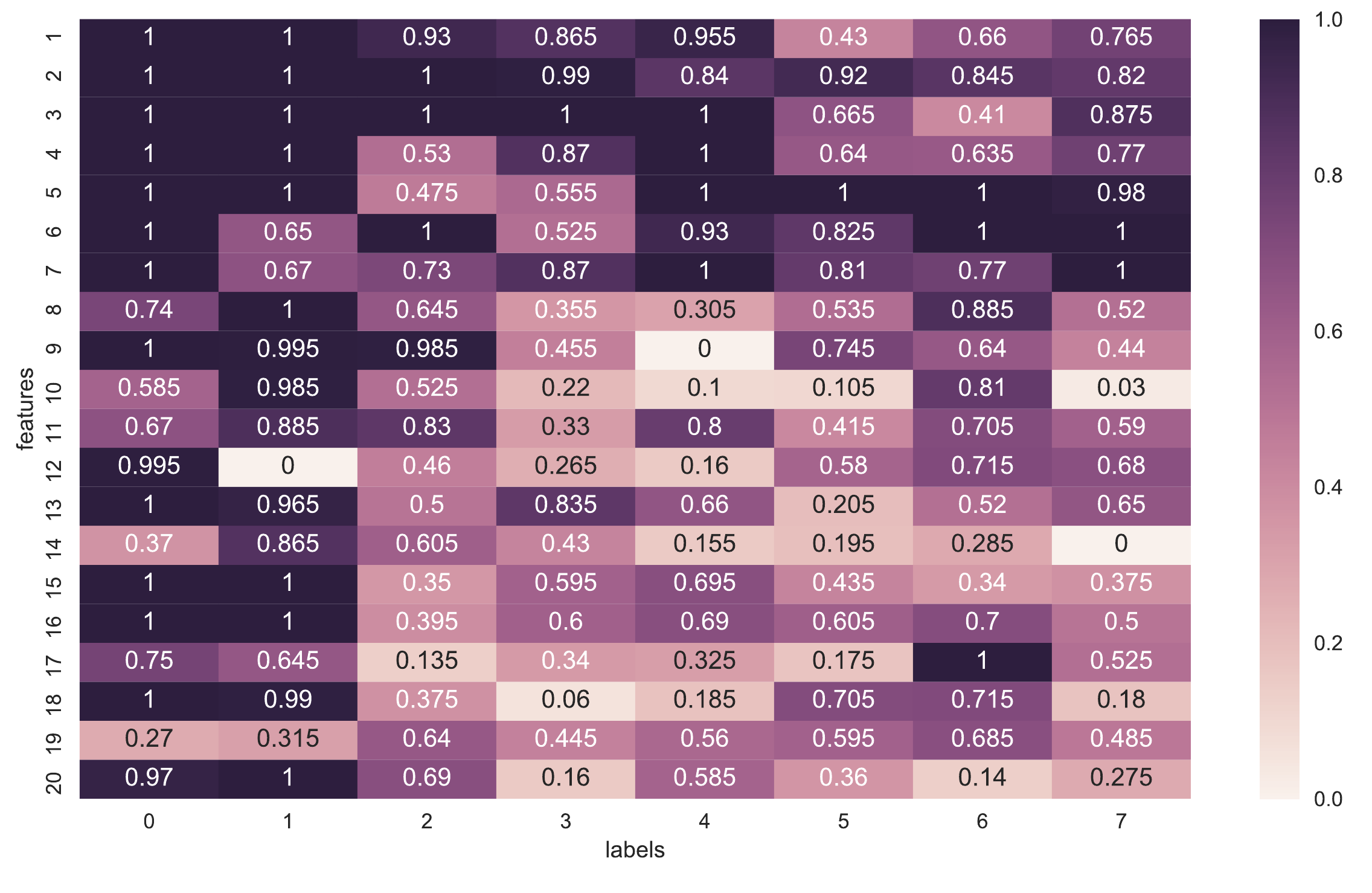}
\vspace*{-1.7em}
\caption{AS-733 Graph. Rows are the node of which ABC value is used as classifier.
Columns are the classes(labels), that is, top 7 nodes being disconnected from
the network one-by-one. Observing top i-th node's (y-axis) ABC value helps
classify the top i-th node's (x-axis) removal.
Class (label) 0 is the graph when there is no change.}
\label{fig:one_feat_annot}
\end{figure*}

To see what is the minimum we need, we tested how much we can identify using
only one feature to train our machine learning algorithm. We use one of the top
20 nodes one by one as feature and train the algorithm with the baseline set of
200 and a set of 200 instances for the removal of top 7 nodes one by one.
Assuming there is no other change in the graph, we store the accuracy achieved
by each feature for each label in a feature - label matrix. Figure~\ref{fig:one_feat_annot} show this matrix. According to the tolerance the
context has, one may say that a $75\%$ accuracy may be enough, or may want at
least $90\%$ accuracy.

\begin{figure*}
\centering
\includegraphics[width = 0.8\textwidth]{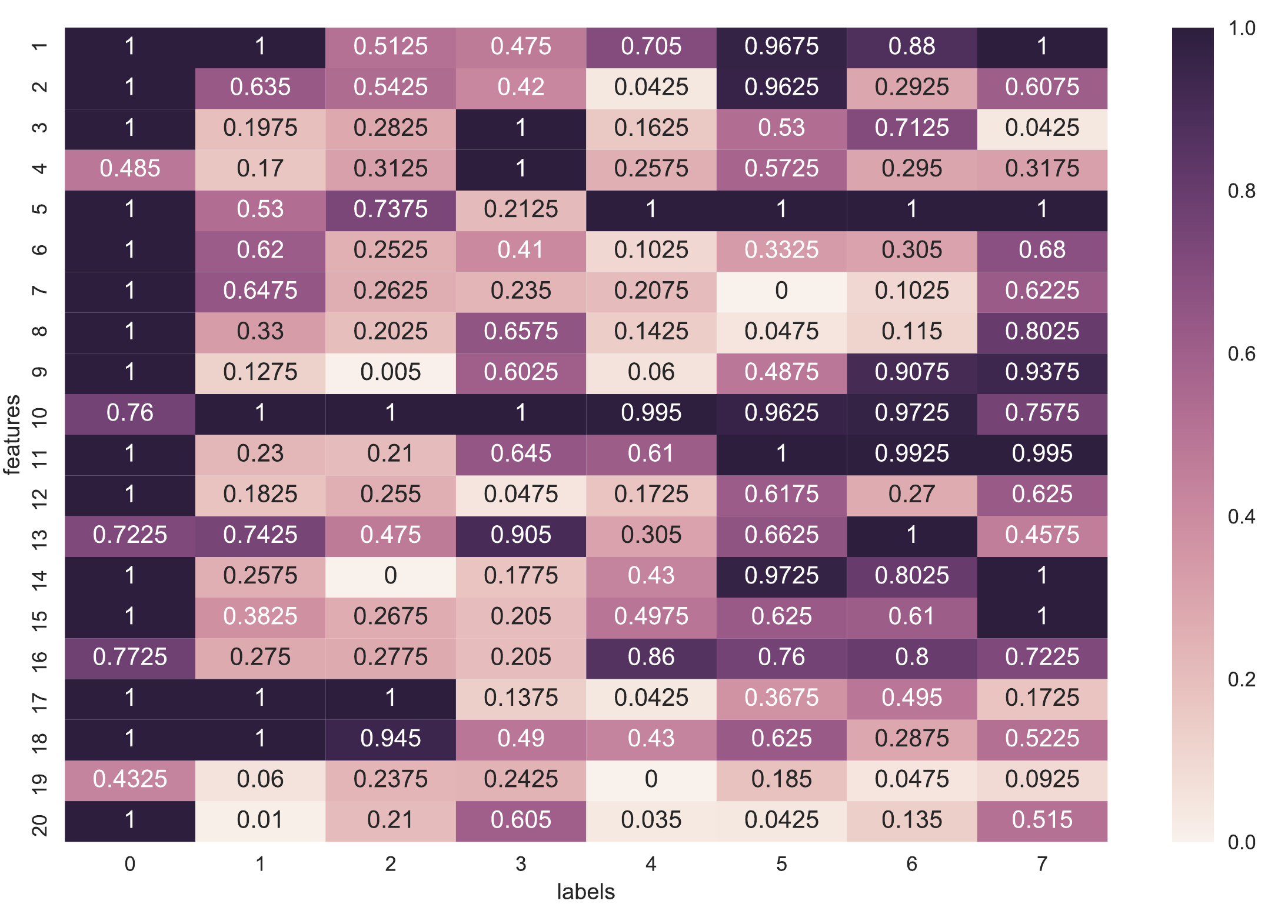}
\vspace*{-1.7em}
\caption{Airports Graph Rows are the node of which ABC value is used as classifier.
Columns are the classes(labels), that is, top 7 nodes being disconnected from
the network one-by-one. Observing top i-th node's (y-axis) ABC value helps
classify the top i-th node's (x-axis) removal.
Class (label) 0 is the graph when there is no change.}
\label{fig:opsahl_onefeat}
\end{figure*}

The next step to further utilize this information would be to find the minimal
number of elements required to successfully detect changes in any of these top
$K$ nodes under the assumption that we do not have any concurrent node failures. This question is also known as minimum set cover problem in
other domains, where we are trying to cover all the columns with the minimal
number of rows selected. For example, in Figure~\ref{fig:one_feat_annot} we
can select $3{rd}$ and $5{th}$ as our features to successfully cover all the
columns (labels/classes). In Figure \ref{fig:opsahl_onefeat}, we can select
$10{th}$ feature alone to have $75\%$ accuracy for first and last labels, and
above $96\%$ for the others. Or, we can select $10{th}$ and $11{th}$ together
to cover all of them with above $96\%$ accuracy.

We also repeated this experiment using not only one but combination of 2 and 3
nodes as the feature set to confirm that using the nodes together would work
and highly probably increase the accuracy compared to using only one feature.
Due a to space limitations, the visual results are not presented.

\subsection{Effect of Noise in Experiments}

It is imperative to consider external/natural factors.
Not every node on the communication graph behaves the same all the time.
We achieve this in our experiments by incorporating noise factor to the
communication probabilities. While each node initiates communication with
others proportional to its initiator (sender) probability, and receives
communication proportional to its receiver probability, these should be
varying from time to time (between intervals) so that we do not assume
that every node behaves exactly the same all the time. For our experiments,
we have added a multiplicative noise factor uniformly random in the  0.8--1.2  range to the communication probability of the nodes.

The cardinality algorithm inherently is not affected by the noise since
that is what we actually want to compute: The active usage and its effective centrality instead of
static structure.

However, the empirical use cases we provided slightly deteriorate in the face
of noisy data. In our experiments, we saw that for the AS-733 and Airports
graphs, the accuracy decreased by 10\% on average and up to 35\% for same
labels in AS-733 graph.

Overall, the algorithms are still able to successfully detect changes in most
cases. More sophisticated detection approaches can be implemented to further
improve quality and make it even more robust to noise, however, this is not in
the scope of this paper.

\section{Conclusion and Discussion}\label{sec:conc}
Detection of critical nodes of a graph has become one of the main starting
points for any graph analysis algorithm. Many importance measures (such as
betweenness and closeness centrality) have been
developed for the purpose of providing  insights for node criticality. These measures  rely on two
assumptions, (i) centrality of a node depends only on the the structure of the
graph and do not take the network activity into account. (ii) the exact graph
structure is known. These assumptions limit the applicability of these methods.
In this work, we have introduced a new metric, active betweenness cardinality,
which can  takes into account network activity while assessing criticality  and  does not  require
knowing of the exact topology.  We showed how this metric can be computed efficiently and how most critical nodes can be identified. We empirically show that the metric can be used to detect significant changes in the topology of the network robustly.

\end{document}